\documentclass[11pt]{article}
\usepackage{epsfig}
\newcommand{\be}{\begin{equation}}
\newcommand{\ee}{\end{equation}}
\newcommand{\ba}{\begin{eqnarray}}
\newcommand{\ea}{\end{eqnarray}}
\def\L{{\cal L}}
\newcommand{\e}{\hbox{e}}
\newcommand{\p}{\partial}
\newcommand{\psibar}{\overline{\psi}}

\newcommand{\vev}[1]{\left\langle #1 \right\rangle}

\newcommand{\ubar}{\overline{u}}
\newcommand{\dbar}{\overline{d}}
\newcommand{\dslash}{\hbox{$\partial$\kern-0.5em\raise0.3ex\hbox{/}}}
\def\slash#1{\hbox{$#1$\kern-0.5em\raise0.3ex\hbox{/}}}

\begin{document}
\begin{titlepage}
\rightline{KOBE-TH-00-02}
\vspace{.5cm}
\begin{center}
{\LARGE Chiral QED out of matter}\\
\renewcommand{\thefootnote}{\fnsymbol{footnote}}
\vspace{1cm} Hidenori SONODA\footnote[2]{E-mail:
sonoda@phys.sci.kobe-u.ac.jp}\\
\renewcommand{\thefootnote}{\arabic{footnote}}
\vspace{.2cm}
Physics Department, Kobe University, Kobe 657-8501, Japan\\
\vspace{.2cm} 
May 2000\\
\vspace{.2cm} 
\end{center}

\abstract In a previous paper we have shown how the Wilsonian
renormalization group naturally leads to the equivalence of the standard
QED with a matter-only theory.  In this paper we give an improved
explanation of the equivalence and discuss, as an example, the
equivalence of a chiral QED in the Higgs phase with a matter-only
theory.  Ignoring the contributions suppressed by the negative powers of
a UV cutoff, the matter-only theory is equivalent to the perturbatively
renormalizable chiral QED with two complex Higgs fields.  In the
matter-only theory chiral anomaly arises without elementary gauge
fields.

\end{titlepage}

\section{Introduction}

The possibility of writing down a matter-only theory which contains all
the interactions of nature was first considered by Heisenberg in his
study of a unified field theory.\cite{H} Along a somewhat different line
of thought, motivated by the model of dynamical symmetry breaking by
Nambu and Jona-Lasinio\cite{NJL}, Bjorken proposed a fermionic theory
without an elementary gauge field which might explain the masslessness
of the photon as a consequence of a symmetry breaking.\cite{Bj} He
actually found his model equivalent to the standard QED, and the
equivalence was further discussed by Bialynicki-Birula, Luri\'e and
Macfarlane, and Guralnik among others.\cite{early} Similarly, the
equivalence of the Nambu-Jona-Lasinio model with a manifestly
renormalizable Yukawa theory was shown by Eguchi.\cite{E}

A clear explanation of such equivalence between a manifestly
renormalizable theory and an apparently non-renormalizable theory was
only relatively recently given in the work of Hasenfratz et
al.\cite{HHJKS} from the viewpoint of the Wilsonian renormalization
group.\cite{WK} It is this viewpoint which was extended to the
equivalence between QED and the Bjorken model in the previous paper by
the author.\cite{S}

The purpose of the present paper is two-fold.  First we improve the
explanation of the equivalence between a manifestly renormalizable model
and an apparently non-renormalizable model.  Using the $1 \over N$
expansions to leading order, we give a simple yet well defined procedure
for building a non-renormalizable model which is equivalent to the
original renormalizable model.  Though the equivalence is guaranteed only
to leading order in $1 \over N$, the possibility of extending the
equivalence beyond the leading order is supported by the general
renormalization group argument.  Second, as a concrete example, we show
how to use the above procedure to construct a matter-only lagrangian
equivalent to a chiral QED which has a much richer structure than the
models of QED discussed in the previous paper.

The paper is organized as follows.  In sect.~2 we give a simpler
explanation of the equivalence of the Bjorken model with QED than was
given in the previous paper.  Using this example, we construct a simple
procedure for constructing a matter-only theory equivalent to the
original model to leading order in $1 \over N$.  In sect.~3 we summarize
the relevant properties of a chiral QED with two flavors of chiral
fermions and two complex scalar fields.  In sect.~4 we follow the
procedure given in sect.~2 to construct a matter-only theory equivalent
to the chiral QED of sect.~3.  We discuss the realization of chiral
anomaly in the matter-only model in sect.~5 before we conclude the paper
in sect.~6.  Three appendices are given for completeness.

We work in the four dimensional euclidean space throughout and use the
same convention for the spinors as in ref.~\cite{S}.

\section{Equivalence revisited}

The equivalence of QED with a matter-only theory was discussed from the
Wilsonian RG (renormalization group) viewpoint in the previous paper
\cite{S}.  In this section we wish to give an improved explanation of
the equivalence.

The standard QED, which is manifestly renormalizable by perturbation
theory, is defined by the lagrangian
\ba
\L_{QED} &=& {1 \over 4 e_0^2} F_{\mu\nu}^2 + {1 \over 2 e_0^2 \xi_0}
(\partial_\mu A_\mu)^2 + {m_0^2 \over 2 e_0^2} A_\mu^2 + {\lambda
\over N (4\pi)^2} {(A_\mu^2)^2 \over 8}\nonumber\\
&& + \psibar^I \left( {1 \over i} \dslash +
{1 \over \sqrt{N}} \slash{A} + i M \right) \psi^I \label{LQED}
\ea
where $I=1,...,N$.  We regularize the theory with a momentum cutoff
$\Lambda$.  To leading order in ${1 \over N}$ the theory is renormalized
by
\ba
{(4 \pi)^2 \over e^2} &\equiv& {4 \over 3} \ln {\Lambda_0^2 \over \mu^2}
= {(4 \pi)^2 \over e_0^2} + {4 \over 3} \ln {\Lambda^2 \over \mu^2} - 1 
\label{landau}\\ {m_\gamma^2 \over e^2} &=& {m_0^2 \over e_0^2} - {2
\over (4\pi)^2} (\Lambda^2 - M^2)\\ {(4\pi)^2 \over e^2 \xi} &=&
{(4\pi)^2 \over e_0^2 \xi_0} + {1 \over 3}
\ea
where $m_\gamma$ is a finite photon mass, and $\mu$ is an arbitrary low
energy renormalization scale.  To satisfy the Ward identity we must
choose
\be
\lambda = - {4 \over 3}
\ee
to leading order in ${1 \over N}$ so that the four-point proper vertex
of the photon field vanishes at zero external momenta.  The momentum
cutoff does not allow the shift of loop momenta, and it does not respect
the gauge invariance of the lagrangian.  Hence, we need to introduce the
self-coupling $\lambda$ to enforce the Ward identity.

Now, let us try to construct a matter-only model which is equivalent to
the above, to leading order in ${1 \over N}$.  There is no unique choice
for the lagrangian, but one straightforward choice is the following:
\be
\L_{Bj} = \psibar^I \left( {1 \over i} \dslash + i M \right)
\psi^I - {1 \over 2N v^2} \left( \psibar^I \gamma_\mu \psi^I \right)^2
+ \Delta \L \label{LBj}
\ee
where
\be
v^2 \equiv {m_0^2 \over e_0^2} = {2 \over (4\pi)^2} (\Lambda^2 - M^2) +
{m_\gamma^2 \over e^2}
\ee
Here, the counterterms $\Delta \L$ are given by
\be
\Delta \L =
{1 \over 4 e_0^2} (\p_\mu B_\nu - \p_\mu B_\mu)^2 + {1
\over 2 e_0^2 \xi_0} (\partial_\mu B_\mu)^2 + {\lambda \over N
(4\pi)^2} {(B_\mu^2)^2 \over 8}
\ee
where
\be
B_\mu \equiv - {(4\pi)^2 \over 2 \Lambda^2} {1 \over \sqrt{N}} \psibar^I
\gamma_\mu \psi^I \label{Bmu}
\ee
We can understand the equivalence between $\L_{QED}$ and $\L_{Bj}$ from
the viewpoint of the Wilsonian RG.\cite{WK} We notice that the
lagrangian (\ref{LBj}) has two distinct critical points.  One is
trivial: $M=0$ and $v$ arbitrary (except for $v_c$ given below).  The
other parameters $e_0^2, \xi_0, \lambda$ are also arbitrary.  This
critical point, which describes $N$ free massless fermions, has
codimension $1$ in theory space, and the fermion mass is the sole
relevant parameter.  There is another non-trivial critical point,
however, which is given by $M=0$ and $v^2 = v_c^2$ (where $v_c^2 = {2
\Lambda^2 \over (4\pi)^2}$ to leading order in $1 \over N$).  The
remaining parameters are arbitrary.  This critical point corresponds to
$N$ free massless fermions and one free massless vector.  The
criticality has codimension $2$ in theory space, and the two relevant
parameters are the fermion and photon masses.  In finding the
non-trivial critical point of the matter-only theory, the $1 \over N$
expansions are helpful.\cite{C} As long as we use perturbation
expansions in powers of the four-fermi interaction, we cannot detect the
second critical point.  The Wilsonian RG tells us that arbitrary
theories which are almost critical are completely characterized by the
relevant and marginally irrelevant parameters.  The corrections are
suppressed by the negative powers of the cutoff.  The critical point of
$\L_{QED}$ at $M=0$, ${m_0^2 \over e_0^2} = v_c^2$ and that of $\L_{Bj}$
at $M=0$, $v^2 = v_c^2$ describe the same criticality; both run to the
same fixed point under the RG.  Therefore, the two lagrangians must
define the same theory.\cite{WK}

The form of the matter-only lagrangian (\ref{LBj}) is by no means
unique.  Any lagrangian in the neighborhood of the non-trivial critical
point will do as long as it has enough degrees of freedom to allow for
two marginally irrelevant parameters $e_0^2, \lambda$.  In ref.~\cite{S}
we have given a different but equivalent lagrangian.

The procedure to get $\L_{Bj}$ out of $\L_{QED}$ can be made systematic. 
We first extract the gaussian part of $\L_{QED}$ quadratic in the gauge
field with no derivatives:
\be
\L_{gauss} = {v^2 \over 2} A_\mu^2 + A_\mu {1 \over \sqrt{N}} \psibar^I
\gamma_\mu \psi^I
\ee
where $v^2 \equiv {m_0^2 \over e_0^2}$.  The equation of motion for this 
lagrangian gives
\be
A_\mu = B'_\mu \equiv - {1 \over \sqrt{N} v^2} \psibar^I \gamma_\mu
\psi^I \label{AB}
\ee
Now we can construct a matter-only lagrangian by substituting the above
interpolating field $B'_\mu$ into $A_\mu$ in the remaining part of the
lagrangian $\L_{QED}$:
\ba
&&\L_{matter} = \psibar^I \left( {1 \over i} \dslash + i M \right)
\psi^I + {v^2 \over 2} \left( A_\mu - B'_\mu \right)^2 - {v^2 \over 2}
{B'_\mu}^2 \nonumber\\
&&+ {1 \over 4 e_0^2} (\p_\mu B'_\nu - \p_\mu B'_\mu)^2 + {1
\over 2 e_0^2 \xi_0} (\partial_\mu B'_\mu)^2 + {\lambda \over N
(4\pi)^2} {({B'_\mu}^2)^2 \over 8}
\ea
Note that the above lagrangian $\L_{matter}$ is quadratic in $A_\mu$
which plays the role of an auxiliary field.  The equation of motion for
$A_\mu$ is given by Eq.~(\ref{AB}), and to leading order in ${1 \over
N}$ we can substitute the equation of motion inside the correlation
functions.  This is because the proper two-point function of $A_\mu$ and
$B'_\nu$ is given by the diagram in Fig.~1 to leading order in ${1 \over
N}$, and the proper part of the correlation function of ${1 \over n!} 
\left({B'_\mu}^2/2\right)^n$ with $2 n$ $A_\mu$ fields is given by $1$
(except for the tensorial factor) to leading order in ${1 \over N}$.
(Fig.~2) The errors of this substitution are suppressed by negative
powers of the cutoff.  Integrating over the auxiliary field $A_\mu$, we
obtain the lagrangian $\L_{Bj}$ (\ref{LBj}).  Except for the mass term
$B_\mu^2$, we can replace $B'_\mu$ by $B_\mu$ since the two differ only
by a normalization which is unity up to ${\mu^2 \over \Lambda^2}$.
Thus, to leading order in ${1 \over N}$, the two lagrangians $\L_{QED}$
(\ref{LQED}) and $\L_{Bj}$ (\ref{LBj}) are equivalent.

\begin{center}
\epsfig{file=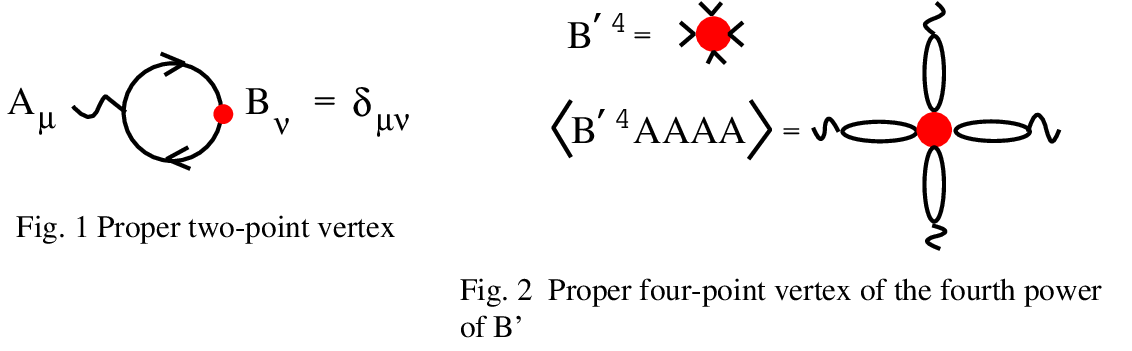, height=3.8cm}
\end{center}
While the equivalence between (\ref{LQED}) and (\ref{LBj}) is strictly
valid only to leading order in ${1 \over N}$, the equivalence can be
generalized beyond the leading order by choosing appropriate values for
the parameters of (\ref{LBj}), as is assured by the general Wilsonian RG
argument.

Finally we note that the lagrangian of the matter-only theory simplifies
somewhat if we raise the momentum cutoff $\Lambda$ to the Landau scale
$\Lambda_0$ defined by (\ref{landau}).  Then the bare charge $e_0$
diverges, and the lagrangian (\ref{LBj}) becomes
\be
\L_{Bj}' = \psibar^I \left( {1 \over i} \dslash + i M \right) \psi^I -
{1 \over 2 N v^2} \left(\psibar^I \gamma_\mu \psi^I\right)^2 + {\lambda
\over N (4\pi)^2} {(B_\mu^2)^2 \over 8}
\ee
This choice of the cutoff is known as the ``compositeness condition'' in
the literature.\cite{early}

We hope that the reader is convinced that any gauge theory (at least as
long as it is abelian) can be rewritten as a matter-only theory.  For
completeness we also derive the equivalence between the manifestly
renormalizable Yukawa model with the Nambu-Jona-Lasinio model along the
above line in Appendix A.

\section{Chiral QED with a momentum cutoff}

We consider a QED with chiral fermions in the Higgs phase.\footnote{We
do not know of any appropriate reference or textbook to point to.}
Contrary to the examples considered in ref.~\cite{S} which had an
explicit photon mass, here the photon mass results dynamically from the
Higgs mechanism.

For our purposes it is important to use a momentum cutoff
regularization.  As we have seen in the previous section, composite
interpolating fields such as $B_\mu$ (\ref{Bmu}) play an important role
in constructing matter-only theories, and their definitions need a
cutoff explicitly.  With a momentum cutoff $\Lambda$, the theory is
defined by the following lagrangian
\ba
&&\L_{chiral} = {1 \over 4e_0^2} F_{\mu\nu}^2 + {1 \over 2 e_0^2 \xi_0}
(\p_\mu A_\mu)^2 + {m_0^2 \over 2 e_0^2} A_\mu^2 \nonumber\\ && +
\ubar^I \left( {1 \over i} \dslash + {1 \over \sqrt{N}} \slash{A} {1 +
\gamma_5 \over 2} \right) u^I + \dbar^I \left( {1 \over i} \dslash - {1
\over \sqrt{N}} \slash{A} {1 + \gamma_5 \over 2} \right) d^I \nonumber\\
&& + {1 \over g_{u,0}^2} \left(\p_\mu - {i \over \sqrt{N}} A_\mu \right)
\phi_u^* \cdot \left( \p_\mu + {i \over \sqrt{N}} A_\mu \right) \phi_u
\nonumber\\ && + {1 \over g_{d,0}^2} \left(\p_\mu + {i \over \sqrt{N}}
A_\mu \right) \phi_d^* \cdot \left( \p_\mu - {i \over \sqrt{N}} A_\mu
\right) \phi_d\nonumber\\ && + {i \over \sqrt{N}} \left( \phi_u
\ubar_R^I u_L^I + \phi_u^* \ubar_L^I u_R^I \right) + {i \over \sqrt{N}}
\left ( \phi_d \dbar_R^I d_L^I + \phi_d^* \dbar_L^I d_R^I \right)\\ && +
{1 \over N} {\lambda_{u,0} \over 4 g_{u,0}^4} \left(\phi_u^*
\phi_u\right)^2 + {M_{u,0}^2 \over g_{u,0}^2} \phi_u^* \phi_u + {1 \over
N} {\lambda_{d,0} \over 4 g_{d,0}^4} \left(\phi_d^* \phi_d\right)^2 +
{M_{d,0}^2 \over g_{d,0}^2} \phi_d^* \phi_d \nonumber\\ && + {1 \over N}
{\tilde{\lambda} \over g_{u,0}^2 g_{d,0}^2} \left(\phi_u^* \phi_u\right)
\left( \phi_d^* \phi_d \right) + \Delta\L \nonumber
\ea
where $\Delta \L$ denotes the counterterms to be given below to
compensate for the gauge non-invariance of the momentum cutoff
$\Lambda$.

The lagrangian is invariant under the following global chiral
transformations:
\ba
u_R^I &\to& \e^{i \theta_u} u_R^I, \quad
\phi_u \to \e^{i \theta_u} \phi_u\\
d_R^I &\to& \e^{i \theta_d} d_R^I,\quad
\phi_d \to \e^{i \theta_d} \phi_d
\ea
The off-diagonal part $\theta_u = - \theta_d$ can be gauged thanks to
the anomaly cancellation between the two flavors $u,d$.  We note that
the above global invariance properties are preserved by the momentum
cutoff regularization.

Though a non-vanishing $\tilde{\lambda}$ is allowed by the above chiral
symmetry, we take $\tilde{\lambda} = 0$ and ignore the mixing between
$\phi_u$ and $\phi_d$ in the rest of the paper.  This is solely to
simplify the calculations.

We now consider the Higgs phase in which both $u$ and $d$ acquire a
mass.  Denoting the fermion mass by $m_i$ for $i=u,d$, we obtain the
following relation to leading order in $1 \over N$:
\be
{1 \over 2} \lambda_{i,0} {m_i^2 \over g_{i,0}^2} + M_{u,0}^2 = {2 \over
(4\pi)^2} g_{i,0}^2 \left(\Lambda^2 - m_i^2 \ln {\Lambda^2 \over
m_i^2}\right)
\ee
Thus, for the theory to be in the Higgs phase we must choose
\be
M_{i,0}^2 < {2 \over (4\pi)^2} g_{i,0}^2 \Lambda^2
\ee
With the following shifts of fields the fields $\rho_i, \varphi_i$ have
vanishing expectation values:
\be
\phi_i = \sqrt{N} m_i + {g_{i,0} \over \sqrt{2}} \left( \rho_i + i
\varphi_i \right)
\ee

To compensate for the non-gauge invariance of the momentum-cutoff
regularization we introduce the following counterterms $\Delta \L$:
\be
\Delta \L = C {1 \over 8 (4\pi)^2 N}
\left(A_\mu^2\right)^2
+ {1 \over (4 \pi)^2 N} A_\mu^2 \left( a_u \phi_u^* \phi_u + a_d
\phi_d^* \phi_d \right)
\ee
To satisfy the Ward identities (which are summarized in Appendix C), we
must make the following choice to leading order in $1 \over N$:
\ba
{m_0^2 \over e_0^2} &=& {2 \Lambda^2 \over (4 \pi)^2} \\
C &=& - {4 \over 3}~, \quad
a_u = a_d  = - {1 \over 2}
\ea
Unlike the case of QED which has the photon mass as a free parameter,
the mass $m_0^2$ is uniquely determined by the Ward identity.

To leading order in ${1 \over N}$ the renormalization is done as
follows:
\ba
e &=& \sqrt{Z_3} ~e_0\\ g_i &=& \sqrt{Z_i} ~g_{i,0} \quad
(i=u,d)\\ {\lambda_i \over 4 g_i^4} &=& {\lambda_{i,0} \over 4 g_{i,0}^4}
+ {1 \over (4 \pi)^2} \ln {\Lambda^2 \over \mu^2}
\ea
where the renormalization constants are defined by
\ba
Z_3 &\equiv& 1 - {4 \over 3} {e^2 \over (4\pi)^2} \ln {\Lambda^2 \over 
\mu^2} = {1 \over {1 + {4 \over 3} {e_0^2 \over (4 \pi)^2} \ln {\Lambda^2
\over \mu^2}}}\\
Z_i &\equiv& 1 - {g_i^2 \over (4 \pi)^2} \ln {\Lambda^2 \over \mu^2} 
= {1 \over 1 + {g_{i,0}^2 \over (4\pi)^2} \ln {\Lambda^2 \over \mu^2}}
\ea
The fermion masses $m_i (i=u,d)$ are unrenormalized to leading order in
$1 \over N$.  The products $e_0 A_\mu$, $g_{i,0} \rho_i$, $g_{i,0}
\varphi_i$ are also left unrenormalized.  We summarize the results for
the renormalized correlation functions in Appendix B.

\section{Matter-only model}

The lagrangian of the matter-only model can be obtained following the
procedure given in sect.~2.  The gaussian part of the lagrangian is
given by
\ba
&&\L_{gauss} = {v^2 \over 2} \left( A_\mu^2 - 2 A_\mu B'_\mu \right)
\nonumber\\ && + V_u^2 \left( \phi_u^* \phi_u - {\Phi'}_u^* \phi_u -
\phi_u^* \Phi'_u \right) + V_d^2 \left( \phi_d^* \phi_d - {\Phi'}_d^*
\phi_d - \phi_d^* \Phi'_d \right)
\ea
where
\ba
v^2 &\equiv& {m_0^2 \over e_0^2} = {2 \over (4\pi)^2} \Lambda^2\\
V_u^2 &\equiv& {M_{u,0}^2 \over g_{u,0}^2} = {2 \over (4\pi)^2}
\Lambda^2 - 2 m_u^2 \left( {\lambda_{u,0} \over 4 g_{u,0}^4} + {1 \over
(4 \pi)^2} \ln {\Lambda^2 \over m_u^2} \right)\\
V_d^2 &\equiv& {M_{d,0}^2 \over g_{d,0}^2} = {2 \over (4\pi)^2}
\Lambda^2 - 2 m_d^2 \left( {\lambda_{d,0} \over 4 g_{d,0}^4} + {1 \over
(4 \pi)^2} \ln {\Lambda^2 \over m_d^2} \right)
\ea
and
\ba
B'_\mu &\equiv& - {1 \over v^2} {1 \over \sqrt{N}} \left( \ubar_R^I
\gamma_\mu u_R^I - \dbar_R^I \gamma_\mu d_R^I \right) \nonumber\\
\Phi'_u &\equiv& - {1 \over V_u^2} {1 \over \sqrt{N}} i \ubar_L^I
u_R^I\quad \Phi_d' \equiv - {1 \over V_d^2} {1 \over \sqrt{N}} i
\dbar_L^I d_R^I 
\ea
Hence, we obtain the following matter-only lagrangian:
\ba
&&\L_{matter} ~=~ \ubar^I {1 \over i} \dslash u^I + \dbar^I {1 \over i}
\dslash d^I - {1 \over 2 N v^2} \left( \ubar_R^I \gamma_\mu u_R^I -
\dbar_R^I \gamma_\mu d_R \right)^2 \nonumber\\ && \quad+ {1 \over N
V_u^2} \left(\ubar_L^I u_R^I\right) \cdot \left( \ubar_R^J u_L^J \right)
+ {1 \over N V_d^2} \left( \dbar_L^I d_R^I \right) \cdot \left(
\dbar_R^J d_L^J \right) + \Delta \L
\ea
The counterterms are given by
\ba
&&\Delta \L = {1 \over e_0^2} {1 \over 4} \left( \p_\mu B_\nu - \p_\nu
B_\mu \right)^2 + {1 \over \xi_0 e_0^2} {1 \over 2} (\p_\mu B_\mu)^2 +
{C \over 8(4\pi)^2 N} \left(B_\mu^2\right)^2\nonumber \\ && \quad + {1
\over g_{u,0}^2} \left( \p_\mu - i {1 \over \sqrt{N}} B_\mu\right)
\Phi_u^* \cdot \left( \p_\mu + i {1 \over \sqrt{N}} B_\mu\right) \Phi_u
\nonumber\\ && \quad + {1 \over g_{d,0}^2} \left( \p_\mu - i {1 \over
\sqrt{N}} B_\mu\right) \Phi_d^* \cdot \left( \p_\mu + i {1 \over
\sqrt{N}} B_\mu\right) \Phi_d \\ && + {B_\mu^2 \over N (4\pi)^2}  
\left( a_u \Phi_u^* \Phi_u + a_d \Phi_d^* \Phi_d \right) + {1
\over N} {\lambda_{u,0} \over 4 g_{u,0}^4} \left( \Phi_u^* \Phi_u
\right)^2 + {1 \over N} {\lambda_{d,0} \over 4 g_{d,0}^4} \left (
\Phi_d^* \Phi_d \right)^2\nonumber
\ea
where
\ba
B_\mu &\equiv& - {(4\pi)^2 \over 2\Lambda^2} {1 \over \sqrt{N}} \left(
\ubar_R^I \gamma_\mu u_R^I - \dbar_R^I \gamma_\mu d_R^I \right)\\ \Phi_u
&\equiv& - {(4\pi)^2 \over 2 \Lambda^2} {1 \over \sqrt{N}} ~i \ubar_L^I
u_R^I, \quad \Phi_d \equiv - {(4\pi)^2 \over 2 \Lambda^2} {1 \over
\sqrt{N}} ~i \dbar_L^I d_R^I
\ea
If we had considered a non-vanishing $\tilde{\lambda}$ in the previous
section, we would have obtained a term proportional to $|\Phi_u|^2
|\Phi_d|^2$ which is allowed by the global symmetry of the lagrangian.

The counterterms $\Delta \L$ are essential for the complete equivalence
to the original chiral QED.  Without $\Delta \L$, the matter-only
lagrangian would depend only on the cutoff $\Lambda$ and the three
parameters $v^2, V_u^2, V_d^2$, and in no way the theory would be
equivalent to the chiral QED which has more number of marginal
parameters.

\section{Chiral anomaly}

The chiral QED has two massless scalar modes $\varphi_u, \varphi_d$, and
one is used for the Higgs mechanism to give the photon a mass, but the
other is left.  The remaining massless mode couples to two photons, and
the renormalized vertex is given by (using the results in Appendix B)
\ba
&&\vev{\left({g_d \over m_d} \varphi_u + {g_u \over m_u}
\varphi_d\right)(-(k+l)) A_\alpha (k) A_\beta (l)} \nonumber\\ && \qquad
\simeq {1 \over \sqrt{2N}} {i \over (4\pi)^2} {8 \over 3} {g_u g_d \over
m_u m_d} \epsilon_{\alpha\beta\mu\nu} k_\mu l_\nu
\ea
Because of the equivalence between the chiral QED and the matter-only
theory of the previous section, the same result is obtained for the
correlation of the corresponding interpolating fields.  In other words
chiral anomaly is correctly reproduced by the matter-only theory.

\section{Conclusion}

The essence of the present and previous papers is that any field theory
can be renormalized by fine tuning the relevant parameters.  This is
what the Wilsonian renormalization group tells us.  Hence, given an
arbitrary lagrangian, whether it is manifestly renormalizable or not, if
we can find a critical point, we get a renormalizable theory by finely
adjusting the relevant parameters so that the physical mass is much
smaller than the cutoff.  The resulting theory depends only on the
relevant and marginal degrees of freedom.  We have elaborated on this
expectation using concrete examples of abelian gauge theories in this
and previous papers.  We have given a very simple procedure for
constructing a matter-only lagrangian which is equivalent to the
original manifestly renormalizable gauge theory.  We have used the $1
\over N$ expansions to locate non-trivial critical points which would be
missed if we used perturbation expansions in non-renormalizable
interactions.

There has been some hope that the theories without elementary gauge
fields or scalar fields would be more tightly constrained.\cite{A} From
the Wilsonian RG viewpoint, this hope is not well founded.\cite{HHJKS}
Unless the symmetry of the theory is enhanced by imposing the vanishing
of a relevant or marginally irrelevant parameter, there is no
justification for the relationship among the free parameters of the
theory.  Writing a matter-only theory in a particular form has, at best,
as much significance as imposing an arbitrary relation among the
parameters of the theory.

The author is aware that much of what is written in this paper may sound
obvious to those versed in the Wilsonian RG.  Since there is no
reference to point to, however, the author hopes that this paper has its
merit in describing a proper and modern way of looking at this old
subject.

\vspace{0.5cm} The author thanks Prof.~K.~Akama for informing him of the
references to early works on the subject.  This work was supported in
part by the Grant-In-Aid for Scientific Research (No.~11640279) from the
Ministry of Education, Science, and Culture, Japan.

\appendix

\section{The Yukawa model vs. the NJL model}

The perturbatively renormalizable Yukawa model, which is invariant under
a chiral $U(1)$, is defined by
\ba
\L_Y &=& \p_\mu \phi^* \p_\mu \phi + {\lambda_0 \over 4 N}
\left( \phi^* \phi \right)^2 + m_0^2 \phi^* \phi \nonumber\\ &&+ \psibar^I
{1 \over i} \slash{\partial} \psi^I + i {g_0 \over \sqrt{N}} \left( \phi~
\psibar_R^I \psi_L^I + \phi^* ~\psibar_L^I \psi_R^I \right)
\ea
with a momentum cutoff $\Lambda$.  By rescaling $\phi$ by ${1 \over g_0}
\phi$, we rewrite the above as
\ba
\L_Y &=& {1 \over g_0^2} \p_\mu \phi^* \p_\mu \phi + {\lambda_0 \over 4
g_0^4 N} \left( \phi^* \phi \right)^2 + {m_0^2 \over g_0^2} \phi^* \phi
\nonumber\\ &&+ \psibar^I {1 \over i} \slash{\partial} \psi^I + i {1
\over \sqrt{N}} \left( \phi~ \psibar_R^I \psi_L^I + \phi^* ~\psibar_L^I
\psi_R^I \right)
\ea
To leading order in ${1 \over N}$, the renormalized parameters are given
by
\ba
{1 \over g^2} &\equiv& {1 \over (4 \pi)^2} \ln {\Lambda_0^2 \over
\mu^2} = {1 \over g_0^2} + {1 \over (4 \pi)^2} \ln {\Lambda^2
\over \mu^2}\\ {m^2 \over g^2} &=& {m_0^2 \over g_0^2} - {2 \over
(4\pi)^2} \Lambda^2\\ {\lambda \over 4 g^4} &=& {\lambda_0 \over 4
g_0^4} + {1 \over (4\pi)^2} \ln {\Lambda^2 \over \mu^2}
\ea
where $\mu$ is an arbitrary low energy renormalization scale.

The theory is in the symmetric (broken) phase if
\be
{m_0^2 \over g_0^2} > \left( < \right) {2 \over (4 \pi)^2} \Lambda^2
\ee

Following the procedure given in sect.~2, we obtain an equivalent
matter-only lagrangian of the generalized Nambu-Jona-Lasinio model as
\ba
\L_{NJL} &=& \psibar^I {1 \over i} \slash{\p} \psi^I + {1 \over N
v^2} \left(\psibar_R^I \psi_L^I\right) \cdot \left( \psibar_L^J \psi_R^J
\right) \nonumber\\ && + {1 \over g_0^2} \p_\mu \Phi^* \p_\mu \Phi + {1
\over N} {\lambda_0 \over 4 g_0^4} \left( \Phi^* \Phi \right)^2 ,
\ea
where $v^2 \equiv {m_0^2 \over g_0^2} = {2 \over (4\pi)^2} \Lambda^2 + {m^2
\over g^2}$, and
\be
\Phi \equiv - {i \over \sqrt{N}} {(4\pi)^2 \over 2 \Lambda^2}
\psibar_L^I \psi_R^I ,\quad
\Phi^* \equiv - {i \over \sqrt{N}} {(4\pi)^2 \over 2 \Lambda^2}
\psibar_R^I \psi_L^I
\ee
If we raise the cutoff $\Lambda$ to the Landau scale $\Lambda_0$, we get
${1 \over g_0^2} \to 0$, and ${\lambda_0 \over 4 g_0^4} \to {\lambda
\over 4 g^4} - {1 \over g^2}$.  Hence, we obtain a simpler lagrangian
\be
\L'_{NJL} = \psibar^I {1 \over i} \slash{\p} \psi^I + {1 \over N
v^2} \left(\psibar_R^I \psi_L^I\right) \cdot \left( \psibar_L^J \psi_R^J
\right) + {1 \over N} \left({\lambda \over 4 g^4} - {1
\over g^2}\right) \left( \Phi^* \Phi \right)^2 
\ee

For an original discussion, see ref.~\cite{HHJKS}.

\section{The renormalized two-, three-, and four-point
functions in chiral QED}

For definiteness we will summarize the results of the lowest order
calculations in $1 \over N$, ignoring the contributions suppressed by
negative powers of the cutoff.  We only list those necessary for
verifying the Ward identities.

\subsection{two-point functions}

For small external momenta, we obtain the following approximate results
for the proper two-point functions:
\ba
\Pi_{A_\mu A_\nu} (k) &\simeq& \left[ {1 \over e^2} + {1 \over (4\pi)^2}
\left({2 \over 3} \ln {\mu^2 \over m_u^2} + {2 \over 3} \ln {\mu^2 \over
m_d^2} - {5 \over 3} \right) \right] \left( k^2 \delta_{\mu\nu} - k_\mu
k_\nu \right) \nonumber\\ && + {1 \over e^2 \xi} k_\mu k_\nu +
\delta_{\mu\nu} ~2 \Bigg[ ~{m_u^2 \over g_u^2} + {m_d^2 \over g_d^2} \\
&& \quad + {1 \over (4 \pi)^2} \left\lbrace m_u^2 \left( \ln {\mu^2
\over m_u^2} - 1\right) + m_d^2 \left( \ln {\mu^2 \over m_d^2} - 1
\right) \right\rbrace \Bigg] \nonumber 
\ea
where $ {1 \over \xi e^2} = {1 \over \xi_0 e_0^2} - {1 \over (4\pi)^2}
{1 \over 3}$.
\ba
\Pi_{\varphi_u \varphi_u} (k) &\simeq& k^2 \left[ 1 + {g_u^2 \over (4
\pi)^2} \left( \ln {\mu^2 \over m_u^2} - 1 \right)\right]\\
\Pi_{\varphi_u A_\nu} (k) &\simeq& i k_\nu \sqrt{2} {m_u \over g_u}
\left[ 1 + {g_u^2 \over (4 \pi)^2} \left(\ln {\mu^2 \over m_u^2} -
1\right)\right]\\ \Pi_{\rho_u \rho_u} (p) &\simeq& p^2 \left[1 + {g_u^2
\over (4\pi)^2} \left ( \ln {\mu^2 \over m_u^2} - {5 \over 3}
\right)\right] \nonumber\\ &&\qquad + \lambda_u {m_u^2 \over g_u^2} + 4
m_u^2 {g_u^2 \over (4 \pi)^2} \left( \ln {\mu^2 \over m_u^2} - 1 \right)
\ea

\subsection{three-point functions}

For small external momenta, we find
\ba
&&\Pi_{\rho_u A_\mu \varphi_u} (p,k,-(p+k)) \simeq {i \over \sqrt{N}}
\Bigg[ (2p+k)_\mu \nonumber\\ &&\qquad + {g_u^2 \over (4\pi)^2}
\left\lbrace k_\mu \left(\ln {\mu^2 \over m_u^2} - 3\right) + p_\mu
\left(2 \ln {\mu^2 \over m_u^2} - 4\right) \right\rbrace \Bigg]\\
&&\Pi_{\rho_u \varphi_u \varphi_u} (p,k,-(p+k)) \simeq - {1 \over \sqrt{2N}}
{g_u \over m_u} \Bigg[ \lambda_u {m_u^2 \over g_u^2}
\nonumber\\ &&\qquad + {g_u^2 \over (4\pi)^2} \left\lbrace 4 m_u^2
\left(\ln {\mu^2 \over m_u^2} - 1\right) - 2 \left(k^2 + kp + {p^2 \over
3}\right) \right\rbrace \Bigg]\\ &&\Pi_{A_\mu A_\nu \rho_u} (k,l,-(k+l))
\nonumber\\ &&\qquad \simeq - {2 \sqrt{2} \over \sqrt{N}} {m_u \over
g_u} \delta_{\mu\nu} \left[ 1 + {g_u^2 \over (4\pi)^2} \left(
\ln {\mu^2 \over m_u^2} - 2\right) \right]\\ &&\Pi_{A_\mu A_\nu \varphi_u}
(k,l,-(k+l)) \simeq {1 \over \sqrt{2N}} {i \over (4\pi)^2} {4 \over 3}
{g_u \over m_u} \epsilon_{\alpha\beta\mu\nu} k_\mu l_\nu
\ea
Note $\Pi_{AAA}$ vanishes due to the CP invariance.

\subsection{four-point functions}

For small external momenta we find
\ba
\Pi_{A_\alpha A_\beta A_\gamma A_\delta} &\simeq& {1 \over N} {1 \over
 (4\pi)^2} {4 \over 3} (\delta_{\alpha\beta} \delta_{\gamma\delta} +
 \hbox{permutations})\\ \Pi_{\varphi_u A_\alpha A_\beta A_\gamma}
 (k,...) &\simeq& {i \over N} {g_u \over m_u} {1 \over (4\pi)^2} {\sqrt{2}
 \over 3} \left( \delta_{\alpha\beta} k_\gamma + \delta_{\alpha \gamma}
 k_\beta + \delta_{\beta\gamma} k_\alpha \right)
\ea

\section{Ward identities}

The following Ward identities must be satisfied.
\ba
&& - i k_\mu \Pi_{A_\mu A_\nu} (k) + \sqrt{2} {m_u \over g_u}
\Pi_{\varphi_u A_\nu} (k) - \sqrt{2} {m_d \over g_d} \Pi_{\varphi_d
A_\nu} (k) = 0\\ && - i k_\mu \Pi_{A_\mu \varphi_u} (k) + \sqrt{2} {m_u
\over g_u} \Pi_{\varphi_u \varphi_u} (k) - \sqrt{2} {m_d \over g_d}
\Pi_{\varphi_d \varphi_d} (k) = 0\\ && - i k_\mu \Pi_{A_\mu \rho_u
\varphi_u}(k,p,-(k+p)) + \sqrt{2} {m_u \over g_u} \Pi_{\varphi_u \rho_u
\varphi_u} (k,p,-(k+p))\nonumber\\ &&\qquad = - {1 \over \sqrt{N}}
\Pi_{\varphi_u \varphi_u} (p+k) + {1 \over \sqrt{N}} \Pi_{\rho_u \rho_u}
(p)\\ && - i k_\mu \Pi_{A_\mu \rho_u A_\nu} (k,p,-(p+k)) + \sqrt{2} {m_u
\over g_u} \Pi_{\varphi_u \rho_u A_\nu} (k,p,-(p+k))\nonumber\\ &&\qquad
= - {1 \over \sqrt{N}} \Pi_{\phi_u A_\nu} (k)\\ && - i (k_\mu + l_\mu)
\Pi_{A_\mu A_\alpha A_\beta} (-(k+l),k,l)\\ && \quad = - \sqrt{2} {m_u
\over g_u} \Pi_{\varphi_u A_\alpha A_\beta} (-(k+l),k,l) + \sqrt{2} {m_d
\over g_d} \Pi_{\varphi_d A_\alpha A_\beta} (-(k+l),k,l)\nonumber\\ && -
i k_\alpha \Pi_{A_\alpha A_\beta A_\gamma A_\delta} (k, ...) \nonumber\\
&& \qquad = - \sqrt{2} {m_u \over g_u} \Pi_{\varphi_u A_\beta A_\gamma
A_\delta} (k,...) + \sqrt{2} {m_d \over g_d} \Pi_{\varphi_d A_\beta
A_\gamma A_\delta} (k,...)
\ea

\end{document}